\documentclass[preprint,12pt]{elsarticle}

\usepackage{amssymb}
\journal{Annals of Physics}

\begin{document}

\begin{frontmatter}



%\fnref{massimo.testa@roma1.infn.it} \fnref{label4}}
%\cortext[cor1]{tel +39 06 49914289}
%\address{\fnref{label3}}
%\fntext[label3]{massimo.testa@roma1.infn.it}
%\fntext[label4]{tel. +39 06 49914289}
\title{The momentum of an electromagnetic wave inside a dielectric}

\author{Massimo Testa \fnref{fn2}}
\ead{massimo.testa@roma1.infn.it}

\address{Dipartimento di Fisica, Universit\`a degli Studi di Roma ``La Sapienza'' \\
   and \\
   INFN -- Sezione di Roma I  \\
   Piazzale A. Moro 2, I-00185 Roma, Italy}
   
\fntext[fn2]{tel. +39 06 49914289}
\begin{abstract}
The problem of assigning a momentum to an electromagnetic wave packet propagating inside an insulator has become known under the name of the Abraham-Minkowski controversy. In the present paper we re-examine this issue making the hypothesis that the forces exerted on an insulator by an electromagnetic field do not distinguish between polarization and free charges. Under this assumption we show that the Abraham expression for the radiation mechanical momentum is highly favoured.
\end{abstract}

\begin{keyword}
Conservation laws \sep dielectric polarisation \sep electromagnetic force on insulators


\end{keyword}

\end{frontmatter}

% \linenumbers

\section{Introduction}

It is well known that an electromagnetic (e.m.) wave propagating in the vacuum carries both energy and momentum\cite{Jack},\cite{landau1}
\begin{eqnarray}
U &=& \int_{{\cal R}_\infty} \frac {1}{2} ({\bf E}^2 + {\bf B}^2) \, d {\bf r} \\
{\bf p} &=& \frac {1}{c} \int_{{\cal R}_\infty} {\bf E} \times {\bf B} \, d {\bf r} \, ,
\end{eqnarray}
where ${\cal R}_\infty$ denotes the three dimensional space.

Also radiation propagating inside matter carries energy and momentum. While the energy is unambiguously given, for an insulating non magnetic material, by\cite{landau}
\begin{equation}
U = \int_{{\cal R}_\infty}{\cal U} \, d {\bf r} \equiv \int_{{\cal R}_\infty} \frac {1}{2} (\epsilon {\bf E}^2 + {\bf B}^2) \, d {\bf r} \, , \label{emene}
\end{equation}
soon after the formulation of Einstein Relativity Theory, conflicting expressions were proposed for the space momentum to be assigned in this situation\cite{Mink},\cite{Abra}. This discrepancy, known as the Abraham-Minkowski controversy, is still attracting a considerable interest in part of the scientific community.\footnote{For a review of the subject see\cite{baxte},\cite{MILO},\cite{kemp} and \cite{griff1}.} The Minkowsky and Abraham proposals for the e.m. momentum in an insulating non magnetic uniform material with dielectric constant $\epsilon$, occupying the full space, are summarized by the equations
\begin{eqnarray}
{\bf p}_M &=& \frac {1}{c} \int_{{\cal R}_\infty} {\bf D} \times {\bf B} \, d {\bf r} = \frac {\epsilon}{c} \int_{{\cal R}_\infty} {\bf E} \times {\bf B} \, d {\bf r} \label{mink11} \\
{\bf p}_A &=& \frac {1}{c} \int_{{\cal R}_\infty} {\bf E} \times {\bf H} \, d {\bf r} = \frac {1}{c} \int_{{\cal R}_\infty} {\bf E} \times {\bf B} \, d {\bf r} \, . \label{abraha11}
\end{eqnarray}
In order to visualize the differences between eqs.(\ref{mink11}) and (\ref{abraha11}), we imagine an idealized situation in which matter is schematized as a uniform, isotropic block of dielectric material ${\cal D}$ with a dielectric constant $\epsilon$, refraction index $n=\sqrt{\epsilon}$, with no magnetic properties, i.e. with permeability $\mu=1$. We will consider a situation in which the matter is sufficiently massive that it can exchange space momentum, but not energy with the incident radiation. The dielectric is surrounded by empty space, ${\cal V}$. If stress variations induced by e.m.\ radiation can be safely neglected, the Abraham-Minkowski discrepancy can be illustrated imagining a collimated e.m.\ wave packet, coming from the vacuum, with an energy $U$ and a momentum ${\bf p}$ ($U =  c |{\bf p}|$), entering the dielectric material. Once the radiation is fully inside the dielectric

\begin{itemize}
\item Minkowski attributes to it the same vacuum energy $U$ and an absolute value of the momentum $|{\bf p}_M| = n |{\bf p}|$,
\item Abraham attributes to it the same $U$ and an absolute value of the momentum $|{\bf p}_A| = \frac {|{\bf p}|}{n}$.

\end{itemize}
From general principles, the momentum change during the time interval $(T_1,T_2)$ is related to the force ${\bf F}_{\cal D} (t)$ that the e.m.\ radiation exerts on the dielectric according to the basic relation
\begin{eqnarray}
\Delta {\bf p} = {\bf p}(T_2) - {\bf p}(T_1) = - \int_{T_1}^{T_2} {\bf F}_{\cal D} (t) dt \, . \label{momforc}
\end{eqnarray}
Eq.(\ref{momforc}) shows that in order to identify the mechanical momentum associated to an e.m.\ wave, we have to know the precise form of the forces exerted on matter by e.m. fields\cite{mans1},\cite{haus},\cite{loud},\cite{mans2},\cite{mans3}. This issue, as many others in Physics, cannot be decided on theoretical grounds alone: the final answer must come from experiments.

It is the purpose of the present paper to test the consequences of the simplest possible hypothesis, ie. that an e.m.\ field acts in the same way on {\em free} and {\em bound} (polarization) charges. Under this assumption we will consider some examples, amenable to quantitative conclusions, which favour the Abraham form of the momentum. Approximate computations are performed in sections \ref{section2}, \ref{section3}, \ref{section4}, assuming that the dielectric susceptibility of the medium is small. In section \ref{sectionnonpert} the problem is treated exactly, studying the crossing of the e.m.\ field through the dielectric boundary in full generality.

\section{Polarization charges and currents}

The description of a transparent, dielectric, homogeneous material with dielectric constant $\epsilon$ and magnetic permeability $\mu =1$ requires the introduction of the polarization field, ${\mathbf P}({\bf x},t)$, in terms of which the polarization charge density $\rho_p({\bf x},t)$ is given by 
\begin{equation}
\rho_p({\bf x},t) =- \nabla \cdot {\bf P}({\bf x},t) \, . \label{VOLUME}
\end{equation}
The time dependence of the polarization field gives rise, as a consequence of charge conservation,  to a volume current density described by
\begin{equation}
{\bf J}_p ({\bf x},t) = {\dot {\bf P}}({\bf x},t) \, ; \label{CURRENT}
\end{equation}
in fact the charge and current densities defined in eqs.(\ref{VOLUME}) and (\ref{CURRENT}) automatically obey the continuity equation
\begin{eqnarray}
\frac {\partial \rho_p}{\partial t} + \nabla \cdot {\bf J}_p = 0 \, .
\end{eqnarray}
The Maxwell equations in the presence of a polarizable medium and in the absence of {\em free} charges and currents  are written as
\begin{eqnarray}
&\nabla \cdot {\bf E} = - \nabla \cdot {\bf P} & \, \, \, \, \, \nabla \cdot {\bf B} = 0 \label{max31} \\
&\nabla \times {\bf E} = - \frac {1}{c} \, {\dot {\bf B}} & \, \, \, \, \,  \nabla \times {\bf B} = \frac {1}{c} \, {\dot {\bf E}} + \frac {1}{c} \, {\dot {\bf P}} \label{max41} \, .
\end{eqnarray}
In the {\em linear} regime, ${\bf P}({\bf x},t)$ is proportional to the electric field ${\bf E}({\bf x},t)$
\begin{equation}
{\bf P}({\bf x},t) = (\epsilon-1) \, {\bf E}({\bf x},t) \equiv \alpha \,\, {\bf E}({\bf x},t) \, , \label{LINEAR}
\end{equation}
which defines the susceptibility $\alpha$. 

In the following we will consider a uniform dielectric, i.e. we will consider the susceptibility $\alpha$ as constant, except in a thin region around the dielectric boundary, in which it  drops quickly to $0$. Whenever possible, without giving rise to ambiguities, we will treat this transition region according to distribution theory\cite{guelf} and will schematize the situation through the introduction of a surface density ${\tilde \sigma}_p$ of polarization charges induced on the dielectric surface $\Sigma$
\begin{equation}
\rho_p({\bf x},t) = - \nabla \cdot {\bf P} ({\bf x},t) =\alpha \, \, ({\bf n}_{\cal D}  \cdot {\bf E}_{\cal D} ({\bf x},t)) \, \delta (\Sigma ({\bf r}))
\equiv {\tilde \sigma}_p ({\bf x},t) \, \delta (\Sigma ({\bf r}))
\, , \label{SURFACE}
\end{equation}
where $ {\bf n}_{\cal D} $ is the outward normal to the dielectric surface, $\Sigma ({\bf r})=0$. $\delta (\Sigma ({\bf r}))$ is the surface Dirac $\delta$-function\cite{guelf} with support on $\Sigma$. In eq.(\ref{SURFACE}) we defined
\begin{eqnarray}
{\tilde \sigma}_p ({\bf x},t) \equiv \alpha \, \, ({\bf n}_{\cal D} \cdot {\bf E}_{\cal D} ({\bf x},t)) \, . \label{SURFACE1}
\end{eqnarray}
The electric field ${\bf E}_{\cal D} ({\bf x},t)$ in eqs.(\ref{SURFACE}) and (\ref{SURFACE1}) is obtained reaching the surface $\Sigma $ from the dielectric side.

According to eq.(\ref{SURFACE}), polarization charges are not present {\em inside} a uniform dielectric
\begin{eqnarray}
&&\rho_p({\bf x},t) =- \nabla \cdot {\bf P}({\bf x},t) = - \alpha \, \nabla \cdot {\bf E}({\bf x},t) =0,\\
&&{\bf x} \notin \Sigma \, , \nonumber
\end{eqnarray}
in virtue of the hypothesis that the sources which generate the incident electromagnetic field are external to the dielectric\cite{mans1}. 

We remind the reader that across the dielectric boundary we have the following discontinuity conditions\cite{Jack}
\begin{eqnarray}
&\Delta E_n = {\tilde \sigma}_p& \, \, \, \, \,  \Delta {\bf E}_{||} = 0 \label{disc1} \\
& \Delta {\bf B} = 0 \, , \label{disc2}
\end{eqnarray}
where $E_n$ and ${\bf E}_{||}$ denote the normal and tangential components of the electric field on $\Sigma$ and the symbol $\Delta$ denotes the variation of the corresponding quantity, while it passes from ${\cal D}$ to ${\cal V}$.

\section{The forces on a dielectric material and the $\alpha$-expansion} \label{section1}

In order to be able to perform explicit computations, we will introduce an approximation scheme. For this purpose we imagine to have an homogeneous, {\it weak} dielectric, namely an insulating material with small susceptibility 
\begin{equation}
\alpha \equiv \epsilon-1 \approx 0 \, . \label{UNO}
\end{equation}
The condition given in eq.(\ref{UNO}) allows us to consider an expansion in powers of $\alpha$, first used by J. P. Gordon\cite{GORDON}.

In the spirit of this approximation scheme we expand the e.m.\ field in powers of $\alpha$ as
\begin{eqnarray}
{\bf E} &\approx& {\bf E}_0 + {\bf E}_1+ \dots \label{ordere}\\
{\bf B} &\approx& {\bf B}_0 + {\bf B}_1+ \dots \, , \label{orderb}
\end{eqnarray}
where the subscript denotes the order in $\alpha$. Thus ${\bf E}_0$ and ${\bf B}_0$ describe the unperturbed e.m.\ field which propagates without feeling any influence from the dielectric material. ${\bf E}_0$ and ${\bf B}_0$ produce in turn a polarization field
\begin{equation}
{\bf P}({\bf x},t) = \alpha \,\, {\bf E}({\bf x},t) \approx \alpha \,\, {\bf E}_0({\bf x},t) \, ,
\end{equation}
which generates, through Maxwell equations, the  first order e.m.\ field ${\bf E}_1$ and ${\bf B}_1$.

We are now in the position to discuss an important feature, common to the Abraham and Minkowski form of the e.m.\ momentum: up to first order in $\alpha$, there is no momentum reflection from the dielectric boundary. We will show this explicitely for the Abraham form of the momentum. Up to first order in $\alpha$ we have, for the Abraham momentum density\footnote{In the Minkowski case we would have ${\bf \Pi}_M \equiv \frac {\epsilon}{c} \, {\bf E} \times {\bf B} \approx (1+\alpha) {\bf \Pi}_0+ {\bf \Pi}_1$.}, eq.(\ref{abraha11}),
\begin{eqnarray}
&&{\bf \Pi} = \frac {1}{c} \, {\bf E} \times {\bf B} \approx \nonumber \\
&&\approx \frac {1}{c} \, {\bf E}_0 \times {\bf B}_0 + \frac {1}{c} \, {\bf E}_1 \times {\bf B}_0 + \frac {1}{c} \, {\bf E}_0 \times {\bf B}_1\equiv {\bf \Pi}_0 + {\bf \Pi}_1 \,,
\end{eqnarray}
where
\begin{eqnarray}
{\bf \Pi}_0 = \frac {1}{c} \, {\bf E}_0 \times {\bf B}_0
\end{eqnarray}
and
\begin{equation}
{\bf \Pi}_1 = \frac {1}{c} \, {\bf E}_1 \times {\bf B}_0 + \frac {1}{c} \, {\bf E}_0 \times {\bf B}_1\, . \label{firstpoyn}
\end{equation}
Eq.(\ref{firstpoyn}) shows that ${\bf \Pi}_1$ can be different from zero only where ${\bf E}_0$ or ${\bf B}_0$ do not vanish and therefore the support of the momentum density ${\bf \Pi} $, coincides with the support of the zeroth order e.m.\ field, ${\bf E}_0$ and ${\bf B}_0$ at all times. Therefore if, at a given time, ${\bf E}_0$ and ${\bf B}_0$ are different from zero only inside the dielectric, the same will be true for ${\bf \Pi}$: in this case no momentum will be present outside the dielectric or, in other words, {\em up to order $\alpha$, momentum is not reflected by the insulator boundary}\footnote{This result is explicitly confirmed in solvable cases, see for instance eq.(9.86) of ref.\cite{griff}.}.

 We now consider a dielectric material invested by an e.m.\ wave, described by a given electric field ${\bf E}_0({\bf x},t)$ and a given magnetic field ${\bf B}_0({\bf x},t)$, generated by sources external to the dielectric itself. 

As discussed in the Introduction, since we deal with infinitely massive matter, it is not possible to trace the momentum flow coming from radiation in a direct way: we can only identify the transferred momentum through the knowledge of the force that the e.m.\ wave exerts on the dielectric, according to eq.(\ref{momforc})\cite{loud1},\cite{loud2}. 

We assume, as discussed in the Introduction, that an e.m.\ field acts on the polarization charges and currents in the same way as it acts on the ${\it free}$ charges and currents\cite{Obukhov}; we will also neglect forces due to a change of the stress status of the dielectric body. 

Under these assumptions the total electromagnetic force acting on the dielectric is given by
\begin{eqnarray}
{\bf F}_{\cal D}(t) = \int_{\cal D} \rho_p ({\bf r},t) \, {\bf E} ({\bf r},t) \, d {\bf r} + \frac {1}{c} \int_{\cal D} {\bf J}_p ({\bf r},t) \times {\bf B} ({\bf r},t) \, d {\bf r} \, . \label{FORCE4}
\end{eqnarray}
If we consider the transition from the dielectric to the vacuum region as sharp, i.e. if we use eq.(\ref{SURFACE}),  eq.(\ref{FORCE4}) becomes
\begin{eqnarray}
{\bf F}_{\cal D}(t) = \int_\Sigma {\tilde \sigma}_p ({\bf r},t) \, {\bf E} ({\bf r},t) \, d \Sigma + \frac {1}{c} \int_{\cal D} {\bf J}_p ({\bf r},t) \times {\bf B} ({\bf r},t) \, d {\bf r} \, . \label{FORCE3}
\end{eqnarray}
Eq.(\ref{FORCE3}) is not well defined because, as shown by eq.(\ref{disc1}), ${\bf E} ({\bf x},t)$ is discontinuous across $\Sigma$ and it is not clear if we should take its value from the vacuum or the insulator side.

We will not, at the moment, worry about this problem, which we shall solve in general in section \ref{sectionnonpert}. Rather we will apply the $\alpha$-approximation scheme and we will show that, up to first order, eq.(\ref{FORCE3}) is well defined.

Using eq.(\ref{SURFACE1}), the force acting on the dielectric ${\cal D}$, formally given by eq.(\ref{FORCE3}), reads, up to order $\alpha$,
\begin{eqnarray}
{\bf F}_{\cal D}(t) \approx \alpha \int_\Sigma ({\bf E}_0 \cdot {\bf n}_{\cal D}) \, \, {\bf E}_0 \, d \Sigma +\frac {\alpha}{c} \int_{\cal D} {\dot {\bf E}_0} \times {\bf B}_0 \, d {\bf r} \label{FORCE}
\end{eqnarray}
and is not affected by any ambiguity, since ${\bf E}_0$ is continuous across $\Sigma$.

The use the identity
\begin{equation}
\frac {1}{c} \, {\bf E}_0 \times {\dot {\bf B}}_0 = ({\bf E}_0 \cdot \nabla ) {\bf E}_0 -\frac{1}{2} \nabla ({\bf E}_0^2) \, ,
\end{equation}
which follows easily from the source-free Maxwell equations, allows us to complete the time derivative in eq.(\ref{FORCE}) and write ${\bf F}_{\cal D}(t)$ in the equivalent form
\begin{eqnarray}
{\bf F}_{\cal D}(t) &\approx& \frac {\alpha}{c} \frac {d}{dt} \int_{\cal D} {\bf E}_0 \times {\bf B}_0 \, d {\bf r} + \frac {\alpha}{2} \int_{\cal D} \nabla ({\bf E}_0^2) \, d {\bf r} = \label{DIECI} \\
&=&  \frac {\alpha}{c} \frac {d}{dt} \int_{\cal D} {\bf E}_0 \times {\bf B}_0 \, d {\bf r} + \frac {\alpha}{2} \int_{\Sigma} {\bf n}_{\cal D} \, \, {\bf E}_0^2 \, \, d\Sigma \, . \label{DIECI1}
\end{eqnarray}
In the following sections we will use eqs.(\ref{DIECI}) and (\ref{DIECI1}) to analyze simple situations.

\subsection{The Einstein box} \label{section2}

In a beautiful paper, Balazs\cite{bala} considered a thought experiment in which an e.m.\ wave packet is sent on a dielectric box. Balazs  used the requirement that the combined center of mass of the box and of the radiation should maintain, during the experiment, a uniform, rectilinear motion. On this basis Balazs concluded that the system goes through three different phases:
\begin{itemize}
\item  while the radiation penetrates into its interior, the box is subject to a force in the direction of the e.m.\ wave;
\item  while the wave is traveling inside the dielectric, the box is not subject to any force;
\item  while the radiation emerges from the dielectric, the box is subject to a force in the opposite direction of the e.m.\ wave.
\end{itemize}
We will analyze the Einstein box problem from the point of view of eq.(\ref{DIECI}), finding results in complete agreement with Balazs analysis and the Abraham form of the momentum\footnote{In this section we follow ref.\cite{GORDON}}.

We imagine ${\cal D}$ as a cube of side $L$, with two faces orthogonal to the $z$-axis, and consider an unperturbed e.m.\ wave packet propagating along the positive $z$-axis, of the form\footnote{In the spirit of the $\alpha$-expansion, ${\bf E}_0({\bf x},t)$ and ${\bf B}_0({\bf x},t)$ are the electric and magnetic fields at zeroth order, $\alpha^0$, and therefore they freely propagate at the speed of light for all $t$.}
\begin{eqnarray}
&&{\bf E}_0({\bf x},t) \equiv (f(z-ct),0,0) \label{free1} \\
&&{\bf B}_0({\bf x},t) \equiv (0, f(z-ct),0) \, , \label{free2}
\end{eqnarray}
where the function $f$ has a finite support
\begin{equation}
f(z-ct) \neq 0 , \,\, \,\, a \leq z-ct \leq b \, . \label{support}
\end{equation}
Eqs.(\ref{free1}) and (\ref{free2}) are valid for $(x,y)$ varying in a transverse cross section with area $A$, smaller than $L^2$, and are supposed to vanish outside it. 

From eq.(\ref{DIECI}) we have
\begin{eqnarray*}
&&{F_{\cal D}}_z(t) \approx \frac {\alpha}{c} \frac {d}{dt} \int_{\cal D} g(z-ct) \, dV +  \frac {\alpha}{2} \int_{\cal D}  \frac {\partial}{\partial z} g(z-ct) \, dV =\\
&&= -  \frac {\alpha}{2} \int_{\cal D} g'(z-ct) \, dV \, ,
\end{eqnarray*}
where $g(z-ct) \equiv f^2(z-ct)$, so that
\begin{equation}
{F_{\cal D}}_z(t) \approx \alpha \, \frac {A}{2} \, [g (-c t) - g(L -c t)] \, . \label{FORCE2}
\end{equation}
If  we consider a situation such that $L > b - a$, then, in virtue of eq.(\ref{support}), $g (-c t)$ and $g(L -c t)$ cannot be both different from $0$ for any $t$ and we can consider three distinct regions of time:
\begin{itemize}
\item Region I, $- \frac {b}{c} \leq t \leq - \frac {a}{c}$, during which the e.m.\ wave is entering in ${\cal D}$ and a force along the wave direction is exerted on the dielectric,
\item Region II, $- \frac {a}{c} \leq t \leq \frac {L- b}{c}$, during which the e.m.\ wave is traveling inside ${\cal D}$ and no force is exerted on the dielectric and
\item Region III, $\frac {L- b}{c} \leq t \leq \frac {L- a}{c}$, during which the e.m.\ wave is exiting from ${\cal D}$ and a force opposite to the wave direction is exerted on the dielectric,
\end{itemize}
in agreement with the Balazs analysis.

\subsection{The e.m.\ momentum inside the Einstein box} \label{section3}

In the Einstein box experiment discussed in section \ref{section2}, the momentum of the incident e.m.\ wave packet is directed along the $z$-axis and its $z$-component is given by
\begin{equation}
{p^{(0)}_\gamma} =  \frac {1}{c} \int_{{\cal R}_\infty} g(z-ct) \, dV =  \frac {A}{c} \int_a^b g(z) dz \, ,
\end{equation}
where ${\cal R}_\infty$ denotes the full three dimensional space.

During the time region I, the momentum transferred to the dielectric block through e.m.\ forces is also directed along the $z$ direction and is given by
\begin{equation}
p_{\cal D}^{(I)} = \int_{- \frac {b}{c}}^{- \frac {a}{c}} {F_{\cal D}}_z (t) dt =  \alpha \, \, \frac {A}{2} \int_{- \frac {b}{c}}^{- \frac {a}{c}} g (-c t) dt = \frac {\alpha}{2}\, \, {p^{(0)}_\gamma} \, .
\end{equation}
Therefore, in order to achieve momentum conservation, we are led to attribute to the e.m.\ wave, while it propagates inside the dielectric during the time region II and not exerting any force on it, a $z$-component of the mechanical momentum such that\footnote{As we have shown in section \ref{section1}, we can safely neglect reflected momentum, while working up to order $\alpha$.}
\begin{equation}
{p^{(II)}_\gamma} = {p^{(0)}_\gamma} - p^{(I)}_{\cal D} = {p^{(0)}_\gamma} (1- \frac {\alpha}{2}) \approx \frac {{p^{(0)}_\gamma}}{n} \, . \label{conserved}
\end{equation}
In eq.(\ref{conserved}) we used the relation
\begin{equation}
n=\sqrt \epsilon \approx 1+\frac {\alpha}{2} \, . \label{index}
\end{equation}
relating the refraction index $n$ to the dielectric constant and in this way we recover the Abraham form of the radiation momentum.

\subsection{The case of oblique incidence} \label{section4}

In this section we discuss a more general kinematic configuration and we show that starting from the expression of the force given in eq.(\ref{DIECI1}) we can also reproduce Snell's law for the refraction of light at the dielectric boundary. 

For this purpose we consider an e.m.\ wave packet incident on a dielectric boundary in an oblique direction. The dielectric occupies the half space $z\geq 0$, always denoted by ${\cal D}$, and, being infinitely massive, it does not exchange energy with the e.m wave hitting it\footnote{This point will be further discussed in section \ref{sectionnonpert}}.

If we integrate eq.(\ref{DIECI1}) over time, from $-\infty$ to $+\infty$\footnote{The $t$-integration is actually limited to the time interval during which the wave packet crosses the boundary at $z=0$.} we get
\begin{eqnarray}
&&{{\bf p}_{\cal D}} = \int_{-\infty}^{+\infty} {\bf F}_{\cal D}(t) dt \approx \nonumber \\
&&\approx \frac {\alpha}{c} ( \int_{\cal D} {\bf E}_0 \times {\bf B}_0 \big |_{t=+\infty} \, d {\bf r} - \int_{\cal D} {\bf E}_0 \times {\bf B}_0\big |_{t=-\infty} \, \, d {\bf r} ) + \nonumber \\
&&+ {\bf n}_{\cal D} \, \frac {\alpha}{2} \int_{-\infty}^{+\infty} dt \int_\Sigma {\bf E}_0^2 \, \, d\Sigma = \nonumber \\
&&=  \frac {\alpha}{c} \int_{\cal D} {\bf E}_0 \times {\bf B}_0 \big |_{t=+\infty} \, d {\bf r} -  {\bf \hat z} \, \frac {\alpha}{2} \int_{-\infty}^{+\infty} dt \int_\Sigma {\bf E}_0^2 \, \, d\Sigma  \label{MOMENTUM} 
\end{eqnarray}
where ${\bf p}_{\cal D} $ is the momentum transferred to the dielectric by the radiation, $\Sigma$ is the dielectric surface, $z=0$, $ {\bf n}_{\cal D}$ its external normal and ${\bf \hat z} =- {\bf n}_{\cal D}$, the unit vector along the positive $z$-axis. Since ${\cal D}$ is the region occupied by the dielectric, the term $\int_{\cal D} {\bf E}_0 \times {\bf B}_0\big |_{t=-\infty} \, \, d {\bf r}$ is zero, because, for sufficiently negative $t$, the radiation is supposed to be confined in the region $z < 0$. Similarly we have 
\begin{equation}
\frac {1}{c} \int_{\cal D} {\bf E}_0 \times {\bf B}_0\big |_{t=+\infty} \, \, d {\bf r} = {\bf p}^{(0)}_\gamma \, , \label{poynting}
\end{equation}
where ${\bf p}^{(0)}_\gamma$ is the {\it initial} e.m.\ momentum; in fact as the time evolves we imagine that the unperturbed e.m.\ radiation will, eventually, be entirely contained inside ${\cal D}$: since the electric and magnetic fields ${\bf E}_0({\bf x},t)$, ${\bf B}_0({\bf x},t)$ propagate freely, the integral appearing in eq.(\ref{poynting}) coincides with the initial momentum of the unperturbed e.m.\ field in the vacuum region.

Summing up we have
\begin{eqnarray}
{\bf p}_{\cal D} = \alpha {\bf p}^{(0)}_\gamma -  {\bf \hat z} \, \,  \frac {\alpha}{2} \int_{-\infty}^{+\infty} dt \int_\Sigma {\bf E}_0^2 \, \, d \Sigma \, . \label{SEMIFINALE}
\end{eqnarray}
In order to give an estimate of the second term on the r.h.s. of eq.(\ref{SEMIFINALE}), we imagine that the initial wave packet predominantly contains short wave lengths and its spatial extension is such that it crosses the dielectric boundary at $z=0$ in a very short time-lapse. In this situation we can neglect the deformation of the e.m.\ wave packet, while it crosses the dielectric boundary. In other words in eq.(\ref{SEMIFINALE}) we can safely assume
\begin{equation}
{\bf E}_0^2 ({\bf x} , t) = D({\bf x} - \, c \, t \, {\bf k}) \, , \label{propagation}
\end{equation}
where ${\bf k}$ is the unit vector which defines the (incident) e.m.\ wave direction.

Performing the change of integration variables ${\bf y} = {\bf x} - \, c \, t \, {\bf k}$, we easily get
\begin{eqnarray}
&&\int_{-\infty}^{+\infty} dt \int_\Sigma {\bf E}_0^2 \, \, d \Sigma = \int_{-\infty}^{+\infty} dt \int_\Sigma D({\bf x} - \, c \, t \, {\bf k}) \, \, d \Sigma = \nonumber \\
&&= \frac {1} {k_z c} \int_{-\infty}^{+\infty} d {\bf y} D({\bf y}) =  \frac {U^{(0)}_\gamma} {c k_z}  =  \frac {p^{(0)}_\gamma} { k_z} \, ,
\end{eqnarray}
where $U^{(0)}_\gamma = c p^{(0)}_\gamma$ is the e.m.\ energy of the free initial wave packet and
\begin{equation}
k_z = \cos {\hat i} \, ,
\end{equation}
where ${\hat i}$ is the angle between the direction of the incident e.m.\ wave and the $z$-axis. In this way we obtain, for the total momentum transferred to the dielectric wall,
\begin{eqnarray}
{\bf p}_{\cal D} = \alpha \, {\bf p}^{(0)}_\gamma -  {\bf \hat z} \, \,  \frac {\alpha}{2 \cos {\hat i} } p^{(0)}_\gamma \, .
\end{eqnarray}
Again, from the absence of reflection and by the requirement of momentum conservation, we are led to attribute to the e.m.\ wave, once inside the dielectric, a momentum ${\bf p}'_\gamma$ such that\begin{equation}
{\bf p}^{(0)}_\gamma={\bf p}'_\gamma + {\bf p}_{\cal D} ={\bf p}'_\gamma +  \alpha \, {\bf p}^{(0)}_\gamma - {\bf \hat z} \, \,  \frac {\alpha}{2 \cos {\hat i} } p^{(0)}_\gamma \, ,
\end{equation}
so that
\begin{eqnarray}
{\bf p}'_\gamma = (1- \alpha) \,  {\bf p}^{(0)}_\gamma + {\bf \hat z} \, \,  \frac {\alpha}{2 \cos {\hat i} } p^{(0)}_\gamma \, .\label{ABRAHAM}
\end{eqnarray}
From eqs.(\ref{ABRAHAM}) and (\ref{index}) we have, to order $\alpha$,
\begin{eqnarray}
&&({\bf p}'_\gamma)^2 = (1- 2 \alpha) \,  ({\bf p}^{(0)}_\gamma )^2 +  \frac {\alpha}{\cos {\hat i} } p^{(0)}_\gamma (p^{(0)}_\gamma)_z=  \nonumber \\
&&= (1- 2 \alpha) \,  ({\bf p}^{(0)}_\gamma )^2 + \alpha ({\bf p}^{(0)}_\gamma )^2 = (1- \alpha) \,  ({\bf p}^{(0)}_\gamma )^2 = \frac {1}{n^2} ({\bf p}^{(0)}_\gamma )^2 \, ,
\end{eqnarray}
which confirms also in this case the Abraham expression for the radiation mechanical momentum.
Moreover eq.(\ref{ABRAHAM}) also implies
\begin{eqnarray}
&&\sin {\hat i}'= \frac {({\bf p}'_\gamma)_\perp}{p'_\gamma} = \frac {(1-\alpha) ({\bf p}^{(0)}_\gamma)_\perp}{(1-\frac {\alpha}{2}) ({p}^{(0)}_\gamma)} = \nonumber \\
&&= (1-\frac {\alpha}{2}) \frac {({\bf p}^{(0)}_\gamma)_\perp}{ ({p}^{(0)}_\gamma)} = \frac {1}{n} \frac {({\bf p}^{(0)}_\gamma)_\perp}{ ({p}^{(0)}_\gamma)}= \frac {1}{n} \sin {\hat i} \, , \label{snell}
\end{eqnarray}
where ${\hat i}'$ is the refraction angle. Eq.(\ref{snell}) shows the validity of Snell's law, up to order $\alpha$, starting from the conservation of the Abraham momentum.

This result is not trivial because Snell's law is usually obtained\cite{Jack} as a consequence of the discontinuity conditions, eqs.(\ref{disc1}) and (\ref{disc2}). In our computation, the e.m.\ field is treated in the zeroth approximation and the discontinuity equations (\ref{disc1}) and (\ref{disc2}) do not appear explicitly in the derivation.

\section{Exact results} \label{sectionnonpert}

Although the small $\alpha$-approximation is useful to get a semi-quantitative idea of the basic phenomena related to dielectric polarization, it is not easy to establish the degree of generality of the results obtained in this way.

In this section we will overcome the limitations due to the $\alpha$-expansion and discuss in general terms the passage of an e.m.\ wave packet from a vacuum region to an insulator.

Following Abraham\footnote{It is straightforward to apply the same procedure to the Minkowski momentum, of course with different results.} we make the hypothesis that the e.m.\ wave momentum in the presence of matter, is given by eq.(\ref{abraha11}). In the following we will trace the variation of the Abraham momentum while the e.m.\ wave crosses the boundary between the empty region ${\cal V}$ and the dielectric ${\cal D}$. In this way we will be able to give a meaning to the formal expression for the force exerted on ${\cal D}$, formally given in eq.(\ref{FORCE3}), obtaining an unambiguous result, which coincides with that found in textbooks\cite{griff} in similar situations.

In order to deal with problems related to the discontinuity of the electric field across $\Sigma$, we introduce a   regularization. In other words we isolate the surface $\Sigma$ enclosing it inside a thin layer ${\cal L}_\delta$ with thickness $\delta$ and will eventually take the limit $\delta \rightarrow 0$, which makes ${\cal L}_\delta$ tend to $\Sigma$.

We denote by ${\cal R}_\infty \ominus {\cal L}_\delta$ the full three dimensional space ${\cal R}_\infty$, deprived of ${\cal L}_\delta$ and consider the quantity
\begin{eqnarray}
{\bf p}_\delta \equiv \frac {1}{c} \int_{{\cal R}_\infty \ominus {\cal L}_\delta} \, \, {\bf E} \times {\bf B} \, d {\bf r} \, . \label{regul1}
\end{eqnarray}
If the integration region were ${\cal R}_\infty$, this expression would coincide with the Abraham momentum of the e.m.\ wave. It is clear, at least from physical intuition, that
\begin{eqnarray}
\lim_{\delta \rightarrow 0} \frac {1}{c}  \int_{{\cal L}_\delta} \, \, {\bf E} \times {\bf B} \, d {\bf r} =0 \, . \label{trapping}
\end{eqnarray}
Eq.(\ref{trapping}) expresses the fact that we do not expect any finite fraction of the e.m.\ momentum to be trapped and accumulated inside ${\cal L}_\delta$ as $\delta \rightarrow 0$. This implies that $\lim_{\delta \rightarrow 0} {\bf p}_\delta$ gives the correct Abraham momentum of the e.m.\ field
\begin{eqnarray}
\lim_{\delta \rightarrow 0} {\bf p}_\delta = {\bf p}_A \, . \label{limabra}
\end{eqnarray}
We are now in the position to compute
\begin{eqnarray}
&&\frac {d  {\bf p}_\delta}{dt}  = \frac {1}{c} \int_{{\cal R}_\infty \ominus {\cal L}_\delta} \, \,{\dot {\bf E}} \times {\bf B} \, d {\bf r} + \frac {1}{c} \int_{{\cal R}_\infty \ominus {\cal L}_\delta} \, \, {\bf E} \times {\dot {\bf B}} \, d {\bf r} \, , \label{ANALOGY}
\end{eqnarray}
where ${\bf E}$ and ${\bf B}$ are to be considered regular since we have excluded the region containing their discontinuities. 

Through the use of the Maxwell eqs.(\ref{max31}) and (\ref{max41}), we can transform some of the volume integrals into surface integrals and get\footnote{We neglect contributions to the surfaces at infinity because we always consider situations in which the e.m.\ field is confined inside a finite region of space.}
\begin{eqnarray}
&&\lim _{\delta \rightarrow 0} \, \frac {d  {\bf p}_\delta }{dt}= \int_{\Sigma} [ - \frac {1}{2} {\bf n}_{\cal D} {\bf E}_{\cal D}^2 + [({\bf n}_{\cal D} \cdot {\bf E}_{\cal D} ) \, {\bf E}_{\cal D}] \, d \Sigma + \{{\cal D} \rightarrow {\cal V} \} + \nonumber \\
&&- \frac {\epsilon -1}{c} \int_{\cal D} {\dot {\bf E}} \times {\bf B} \, d {\bf r} \, , \nonumber
\end{eqnarray}
where ${\bf E}_{\cal D}$ and ${\bf n}_{\cal D}$ are the electric field and the external normal respectively, computed on $\Sigma$ as a limit from the dielectric side. $\{ {\cal D} \rightarrow {\cal V} \}$ represents the same expression, computed as a limit from the vacuum side. There is no term containing the magnetic field, as a consequence of its continuity across $\Sigma$, eq.(\ref{disc2}).

Using eqs.(\ref{disc1}) and (\ref{limabra}) we get
\begin{eqnarray}
&&\lim _{\delta \rightarrow 0} \frac {d {\bf p}_\delta }{dt} =  \frac {d {\bf p}_A} {dt}  = \int_{\Sigma} [ - \frac {1}{2} {\bf n}_{\cal D} \Delta {\bf E}^2 + \Delta (E_n \, {\bf E})] \, d \Sigma - \frac {\epsilon -1}{c} \int_{\cal D} {\dot {\bf E}} \times {\bf B} \, d {\bf r} = \nonumber \\
&&= - \frac {1}{2} \int_{\Sigma} {\tilde \sigma}_p \, ( {\bf E}_{\cal {\cal V}} + {\bf E}_{\cal D}) d \Sigma - \frac {1}{c} \int_{\cal D}  {\bf J}_p \times {\bf B} \, d {\bf r} \equiv - {\tilde {\bf F}}_{\cal D}(t) \, , \label{poyntgen}
\end{eqnarray}
where
\begin{eqnarray}
&&{\tilde {\bf F}}_{\cal D}(t) = \frac {1}{2} \int_{\Sigma} {\tilde \sigma}_p \, ( {\bf E}_{\cal V}  + {\bf E}_{\cal D}) d \Sigma + \frac {1}{c} \int_{\cal D}  {\bf J}_p \times {\bf B} \, d {\bf r} = \label{poyntgen1} \\
&&=\frac {1}{2} \int_{\Sigma} {\tilde \sigma}_p \, ( {\bf E}_{\cal V}  + {\bf E}_{\cal D}) d \Sigma + \frac {\epsilon -1}{c} \int_{\cal D} {\dot {\bf E}} \times {\bf B} \, d {\bf r} \, . \label{poyntgen2}
\end{eqnarray}
On the other hand, denoting by ${\bf p}_{\cal D}$ the dielectric momentum, from total momentum conservation we have
\begin{eqnarray}
\frac {d }{dt} {\bf p}_{\cal D} = - \frac {d }{dt} {\bf p}_A = {\tilde {\bf F}}_{\cal D}(t) \, . \label{forceabr}
\end{eqnarray}
Eq.(\ref{forceabr}) makes the expression ${\tilde {\bf F}}_{\cal D}(t)$ in eq.(\ref{poyntgen1}) a very natural candidate for the force exerted by the e.m.\ field on an insulator, once we assume the Abraham expression for the momentum. Eq.(\ref{poyntgen1}) gives the correct prescription to be used in cases when the discontinuity of the electric field cannot be neglected: the force must be computed using the average of the electric field on $\Sigma$ from the vacuum and from the dielectric side. This is the natural choice when we compute the electrostatic force acting on a simple layer charge distribution, as discussed in textbooks\cite{griff}. With this prescription eq.(\ref{poyntgen1}) gives a complete formalization of the statement that polarization charges and currents interact in an universal way with the e.m.\ field.

A similar computation, performed with the regularized Minkowski momentum eq.(\ref{mink11}),
\begin{eqnarray}
{{\bf p}_M}_\delta \equiv \frac {1}{c} \int_{{\cal R}_\infty \ominus {\cal L}_\delta} \, \, {\bf E} \times {\bf B} \, d {\bf r}+\frac{\alpha}{c} \int_{\cal D} \, \, {\bf E} \times {\bf B} \, d {\bf r}
\label{regul12}
\end{eqnarray}
would give a result very different from that of eqs.(\ref{poyntgen1}) and (\ref{poyntgen2}) and completely unnatural.

As a consistency check we notice that ${\tilde {\bf F}_{\cal D}} = 0 $ as soon as the e.m.\ field does not touch the boundary of ${\cal D}$. This is obvious for the first term of eq.(\ref{poyntgen2}). As for the second term, $ \frac {\epsilon -1}{c} \int_{\cal D} {\dot {\bf E}} \times {\bf B} \, d {\bf r}$, when ${\bf E}$ and ${\bf B}$ are zero on the boundary $\Sigma$, so that $\epsilon$ discontinuities are not present in the region of integration, it can be transformed through partial integration, using Maxwell equations, as
\begin{eqnarray}
&& \frac {\epsilon -1}{c} \int_{\cal D} {\dot {\bf E}} \times {\bf B} \, d {\bf r} =  \frac {\epsilon-1}{\epsilon} \, \int_{\cal D} ( \nabla \times {\bf B}) \times {\bf B} \, d {\bf r} = \\
&&=  \frac {\epsilon-1}{\epsilon} \int_\Sigma [ ({\bf n}_{\cal D} \cdot {\bf B}) \,  {\bf B} - \frac{1}{2} {\bf B}^2 \, {\bf n}_{\cal D}] \, d {\bf r} \, .
\end{eqnarray}
Therefore, as soon as the e.m.\ wave leaves $\Sigma$, it ceases to exert any force on the dielectric, in agreement with the result of the approximate computation of section \ref{section2}.

Through arguments similar to those discussed above, we can also study the energy flow of the e.m.\ field while it crosses the dielectric boundary.

The e.m.\ energy, $U$, in the presence of a dielectric material is given by eq.(\ref{emene}), which, in analogy with eq.(\ref{regul1}), can be regularized as
\begin{eqnarray}
U_\delta \equiv \frac {1}{2} \int_{{\cal R}_\infty \ominus {\cal L}_\delta} \, \, (\epsilon {\bf E}^2 + {\bf B}^2) \, d {\bf r} \, .
\end{eqnarray}
We can compute
\begin{eqnarray}
&&\frac {d U_\delta}{dt}  =  \int_{{\cal R}_\infty \ominus {\cal L}_\delta} \, \, (\epsilon {\bf E} \cdot {\dot {\bf E}} + {\bf B} \cdot {\dot {\bf B}}) \, d {\bf r} = \nonumber \\
&&= - \int_\Sigma [ ({\bf E}_{\cal D} \times {\bf B}_{\cal D}) \cdot {\bf n}_{\cal D} +  ({\bf E}_{\cal V} \times {\bf B}_{\cal V}) \cdot {\bf n}_{\cal V}] \, d \Sigma = \nonumber \\
&&= - \int_\Sigma \Delta [ ({\bf E} \times {\bf B} ) \cdot {\bf n}_{\cal D} ] \, d \Sigma \, . \label{enerflow}
\end{eqnarray}
As a consequence of eqs.(\ref{disc1}) and (\ref{disc2}) we find
\begin{eqnarray}
 \Delta [ ({\bf E} \times {\bf B} ) \cdot {\bf n}_{\cal D}] = ( \Delta {\bf E} \times {\bf B} ) \cdot {\bf n}_{\cal D} = {\tilde \sigma} ( {\bf n}_{\cal D} \times {\bf B} ) \cdot {\bf n}_{\cal D} = 0 \, ,
\end{eqnarray}
so that eq.(\ref{enerflow}) gives
\begin{eqnarray}
\frac {d U}{dt} \equiv  \lim_{\delta \rightarrow 0} \frac {d U_\delta}{dt}  = 0 \, ,
\end{eqnarray}
which shows that there are no energy radiation transfers in the transition from ${\cal V}$ to ${\cal D}$.

\section{Conclusions}

In this paper we addressed the problem of the interaction between an e.m.\ field and an ideal insulator. Although ultimately the answer to this kind of questions has to come from experiments, we explored the consequences of what looks like the most natural assumption, namely that the e.m\ field acts in the same way on free and on polarization charges. Within this framework we presented in sections \ref{section2}, \ref{section3} and \ref{section4}, through the use of the $\alpha$-expansion, some considerations which support a strict relation between this hypothesis on the forces and the validity of the Abraham form for the radiation mechanical momentum. In particular we were able to connect, up to order $\alpha$, Snell's refraction law to the Abraham momentum conservation.

Finally, in section \ref{sectionnonpert}, we presented a general argument, not requiring the $\alpha$-expansion, which also clarified the correct way of dealing with the discontinuity of the electric field across a dielectric boundary.

\section*{Acknowledgements}
It is a pleasure to thank professor Giancarlo Ruocco for introducing me to the problems treated in this paper and professor Giancarlo Rossi for discussions on these subjects.

I also want to thank the Referee for asking a stimulating question on the arguments presented in section \ref{sectionnonpert}, which resulted into a better understanding of the subject.

\end{document}